\documentclass[aps,pra,preprint,showpacs,nofootinbib,floatfix]{revtex4}
\usepackage{amsmath,dcolumn,graphicx,longtable,epstopdf}
\begin{document}
\title{Long-range forces between two excited mercury atoms and associative ionization}
\author{James S. Cohen}
\email{cohen@lanl.gov}
\affiliation{Theoretical Division, Los Alamos National Laboratory,
Los Alamos, New Mexico 87545}
\author{Andrei Derevianko}
\email{andrei@unr.edu} \affiliation{Department of Physics,
University of Nevada, Reno, Nevada 89557}
\date{\today}
\begin{abstract}
The long-range quadrupole-quadrupole ($\sim R^{-5}$) and leading
dispersion ($\sim R^{-6}$) interactions between all pairs of excited
Hg($6s6p$) $^3P_0$, $^3P_1$, $^3P_2$, and $^1P_1$ atoms are determined.
The quadrupole moments are calculated using the {\it ab initio} relativistic
configuration-interaction method coupled with many-body perturbation
theory.  The van der Waals coefficients are approximated using
previously calculated static polarizabilities and expressions for
the dispersion energy that are validated with similar systems.
The long-range interactions are critical for associative ionization in
thermal and cold collisions, and are found to be quite different for
different pairs of interacting states.  Based on this knowledge and the
short-range parts of previously calculated potential curves,
improved estimates of the chemi-ionization cross sections are obtained.
\end{abstract}
\maketitle

\section{INTRODUCTION}\label{sec:introduction}
Potential-energy curves for the interaction of two mercury atoms in their lowest excited configuration ($6s6p$) have previously been calculated and used to estimate the chemi-ionization cross sections for collisions of two such atoms \cite{cohen02}.  However, those potential curves were not accurate at distances beyond the valence region, which are of great importance in thermal and cold collisions.  Reactions at such energies are pertinent to modeling of fluorescent lamps as well as recent considerations of mercury as a candidate for an atomic clock \cite{AtomicClock}.  In the present work, the long-range interactions are calculated and combined with the short-range parts of the previously calculated potential curves \cite{cohen02} in order to provide better estimates of the chemi-ionization cross sections.

Chemi-ionization is allowed if the potential energy of the two interacting Hg atoms lies higher than the potential energy of ground-state Hg$_2^+$ at any internuclear distance $R$.  For Hg($^3P_0$)+Hg($^3P_0$), Hg($^3P_1$)+Hg($^3P_0$),
Hg($^3P_1$)+Hg($^3P_1$), Hg($^3P_2$)+Hg($^3P_0$), and
Hg($^3P_2$)+Hg($^3P_1)$ only associative ionization
\begin{equation}
{\rm Hg}^*(6s6p) + {\rm Hg}^*(6s6p) \rightarrow
{\rm Hg}_2^+ + e^-  ,
\end{equation}
utilizing the attractive interaction energy between Hg$^+$ and ground-state Hg,
is possible at very low energies since their total excitation energies are smaller than the ionization potential of Hg.
However, as discussed in Ref.\ \cite{cohen02}, based on the potential curves of Hg$_2^*$ and Hg$_2^+$, essentially all chemi-ionization in mercury at very low collision energies, even for the other reactant pairs, Hg($^3P_2$)+Hg($^3P_2$),
Hg($^1P_1$)+Hg($^3P_0$), Hg($^1P_1$)+Hg($^3P_1$), Hg($^3P_2$)+Hg($^1P_1$), and Hg($^1P_1$)+Hg($^1P_1$), can be expected to be of the associative type and will be referred to as associative ionization (AI) henceforth.

For nonsymmetric atomic states (i.e., both atoms having $J\ne 0$), the leading term in
the long-range interaction is generally of the
form $C_5/R^5$, where $C_5$ can be positive or negative.  However, we will show that $J\ne 0$ is neither a necessary nor sufficient condition for $C_5\ne 0$.  If $C_5$ is
positive and sufficiently large, then reactions such as AI, which
occur at short range, are precluded.  In all cases there is also an attractive van
der Waals interaction of the form $C_6/R^6$.

For the ten possible combinations of the $^3P_0$,  $^3P_1$,
$^3P_2$, and  $^1P_1$ states that may react, there are total of 90
potential-energy curves describing the interactions.  The AI cross
section of only one of these pairs---Hg($^3P_1$)+Hg($^3P_0$)---has
been measured experimentally (although one experiment apparently
misidentified these reactants as Hg($^3P_0$)+Hg($^3P_0$)).  This
experimental value is particularly useful in validating the present
cross sections since the autoionization width has not yet been
calculated but is assumed to be large.  The experimental cross
section, combined with the theoretical potential curves, suggests
that the autoionization width is indeed sufficiently large to
saturate ionization when the continuum is penetrated at short range.

\section{Quadrupole moments of atoms and molecular $C_5$ coefficients}\label{sec:quadrupole}

The quadrupole moments of the mercury atom in its $6s6p$ $^3P_1$,  $^3P_2$,  and $^1P_1$ states (the quadrupole moment of the $^3P_0$ atom vanishes by symmetry), as well as the off-diagonal matrix elements, are calculated using the {\it ab initio} relativistic valence configuration-interaction (CI) method based on the Brueckner orbitals \cite{BruecknerOrbitals}, which effectively includes core polarization.  In addition, the random-phase-approximation (RPA) chain of diagrams is included.  This method expands the two-particle basis functions as
\begin{equation}
\Psi(\pi,J,M)=\sum_{k\ge l}c_{k,l}\Phi_{k,l}(\pi,J,M) ,
\end{equation}
where $\pi$ is the parity of the state $\Psi$ and $J$ and $M$ are the
total angular momentum and its projection.
The weights  $c_{k,l}$ are determined by solving the eigenvalue
problem based on the Hamiltonian in the model space spanned
by
the basis functions defined in the subspace of virtual orbitals
\begin{equation}
\Phi_{k,l}(\pi,J,M)=
\eta_{k,l}\sum_{m_k,m_l}c_{j_k,m_k;j_l,m_l}^{J,M}
a_{\{n_k,\kappa_k,m_k\}}^{\dagger}a_{\{n_l,\kappa_l,m_l\}}^{\dagger}
|0_{\rm core}\rangle \,.
\end{equation}
Here the sets $\{n,\kappa,m\}$ enumerate quantum numbers, the $\eta_{k,l}$
are normalization factors, the $a^{\dagger}$ are creation operators,
and the quasivacuum state $|0_{\rm core}\rangle$ corresponds
to a closed-shell core containing 78 of the 80 electrons of the Hg atom.
The one-particle orbitals $\phi_k$ are determined by solving a
Dirac equation
\begin{equation}
 \left( h_0 + V_\mathrm{DHF} + \Sigma \right) \phi_k = \varepsilon_k \phi_k \, .
\end{equation}
Here $h_0$ includes the rest mass term, kinetic energy,
and the Coulomb interaction with the nucleus. $V_\mathrm{DHF}$ is the
Dirac-Hartree-Fock (DHF) potential due to
core electrons. $\Sigma$ is the self-energy operator computed in the
second-order approximation (see, e.g., Ref.~\cite{SavJoh02}).
Qualitatively, this correction describes a response
of a valence electron to polarization of the core by the electron's own field and is the dominant
correlation effect. The resulting
one-particle orbitals are usually referred to as Brueckner orbitals. Once the
Brueckner orbitals
are computed, we solve the CI problem in the model space.
The model space Hamiltonian includes the Coulomb interaction between
valence electrons.  We do not include so-called screening corrections to
the Hamiltonian. As demonstrated~\cite{SavJoh02}, these corrections,
although computationally expensive, are relatively insignificant for divalent atoms.

With the determined CI wave functions we form a matrix element of the quadrupole
operator. The associated one-particle reduced quadrupole matrix elements are given by
\begin{equation}\label{eq: RedMatEl}
\langle\phi_i || Q || \phi_j\rangle = \langle\kappa_i || C^{(2)} || \kappa_j \rangle \int_0^{\infty} r^2[G_i(r)G_j(r)+F_i(r)F_j(r)]dr ,
\end{equation}
where $C^{(2)}$ is the normalized spherical harmonic \cite{varshalovich88},
and $G$ and $F$ are the large and small radial components of the relativistic
wave function.  Further we employ the RPA approximation \cite{amusia75}, which
accounts for shielding of an externally applied field by the core electrons
and substitutes dressed matrix elements for the bare one-particle matrix elements.

The quadrupole-quadrupole interaction energy between two atoms is  given by \cite{knipp38, chang67, derevianko01}
\begin{equation}
V_{QQ}=\frac{1}{R^5}\sum_{\mu=-2}^{2}w(\mu) (Q_{\mu})_{I}(Q_{-\mu})_{II},
\end{equation}
where
\begin{equation}
w(\mu)=\frac{4!}{(2-\mu)!(2+\mu)!} .
\end{equation}
The quadrupole spherical tensor is defined by\footnote{Atomic units are used except where explicitly indicated otherwise. }
\begin{equation}
Q_{\mu}=-\sum_i r_i^2 C_{\mu}^{(2)}(\hat{\bf r}_i)
\end{equation}
where the sum goes over all atomic electrons (though in the case of Hg($6s6p$) only the unpaired $6p$ electron contributes).  The quadrupole moments ${\cal Q}(^{2S+1}P_J$) are defined, as conventional, by the component in the ``stretched'' $M=J$ state:
\begin{equation}
{\cal Q}(^{2S+1}P_J)=2\langle ^{2S+1}P_{J, J}\ |Q_0| \ ^{2S+1}P_{J, J} \rangle
\end{equation}
(note the factor of 2; some definitions don't include this factor \cite{landau65}) and is related to the reduced matrix element, given by Eq.\ (\ref{eq: RedMatEl}), by a factor derived from the Wigner-Eckart theorem \cite{weissbluth78},
\begin{equation}
c_{\rm WE}(J)=\left( \begin{array}{ccc}J & 2 & J \\ -J & 0 & J \end{array} \right) ,
\end{equation}
so that
\begin{equation}
{\cal Q}(^{2S+1}P_J)=2c_{\rm WE}(J) \langle ^{2S+1}P_J \;||Q||\; ^{2S+1}P_J \rangle .
\end{equation}
Both the reduced matrix elements and the conventional quadrupole moments are given in Table \ref{table: quadrupole}.
The accuracy of these moments is expected to be a few percent (for example,
we find that the RPA  shielding reduces the CI values for quadrupole
moments by less than 3\%).
We also note that nonrelativistically the ratio
${\cal Q}(^3P_2) /{\cal Q}(^3P_1) = -2$, while in our
calculations this ratio is  $-2.45$. This deviation reflects the importance
of relativistic corrections for the heavy mercury atom.

For a particular pair of atomic states $^{2S_a+1}P_{J_a}$ and $^{2S_b+1}P_{J_b}$, the Hamiltonian $H$ for the reduced quadrupole moment can be obtained in a double atomic basis.  For two atoms in the same state, $\psi=|S_a,L_a,J_a,M_a\rangle |S_a,L_a,J_a,M_b\rangle$ and $\psi^{\prime}=|S_a,L_a,J_a,M_a^{\prime}\rangle |S_a,L_a,J_a,M_b^{\prime}\rangle$ for the initial and final state, respectively.  For the two  like atoms in different states, $(S_a,L_a,J_a) \neq  (S_b,L_b,J_b)$, symmetric and antisymmetric linear combinations  of the functions on centers I and II must be formed,
\begin{equation}
\psi = 2^{-1/2}(| S_a,L_a,J_a, M_a \rangle_{I} |S_b,L_b, J_b ,M_b \rangle_{II} \pm | S_a,L_a,J_a, M_a \rangle_{II} | S_b,L_b,J_b ,M_b \rangle_{I})
\end{equation} 
and
\begin{equation}
\psi^{\prime} = 2^{-1/2}(| S_a,L_a,J_a, M_a^{\prime} \rangle_{I} | S_b,L_b, J_b ,M_b^{\prime} \rangle_{II} \pm | S_a,L_a,J_a, M_a^{\prime} \rangle_{II} | S_b,L_b, J_b ,M_b^{\prime} \rangle_{I}) .
\end{equation} 
The quadrupole--quadrupole interaction is then given by
{\setlength{\mathindent}{-1.0mm} \setlength{\arraycolsep}{1.0mm}
\begin{eqnarray} \label{eq:H1}
&&\langle \psi | Q_{\mu}Q_{-\mu} | \psi^{\prime}\rangle =  R^{-5}   \nonumber  \\ 
&& \times \left[ \sum_{\mu} w(\mu) (-1)^{J_a-M_a}   (-1)^{J_b-M_b} 
 \left(\! \begin{array}{ccc}J_a & 2 & J_a \\ -M_a & \mu & M_a^{\prime} \end{array} \!\right)
\left(\! \begin{array}{ccc}J_b & 2 & J_b \\ -M_b & -\mu & M_b^{\prime} \end{array} \!\right) \right.
 \langle a||Q||a \rangle  \langle b||Q||b \rangle \nonumber \\ 
 && \pm (1-\delta_{a,b}) \left. \sum_{\mu} w(\mu) (-1)^{J_a-M_a}   (-1)^{J_b-M_b}  
 \left(\! \begin{array}{ccc}J_a & 2 & J_b \\ -M_a & \mu & M_b^{\prime} \end{array} \!\right)
\left(\! \begin{array}{ccc}J_b & 2 & J_a \\ -M_b & -\mu & M_a^{\prime} \end{array} \!\right) 
\langle a || Q || b \rangle  \langle b || Q || a \rangle \right] \nonumber  \\ 
\end{eqnarray} 
\!} 
in terms of the Wigner 3-j symbols and the reduced matrix elements, which satisfy
$\langle a || Q || b \rangle = (-1)^{J_a-J_b}\langle b || Q || a \rangle$. 
By the properties of the 3-j symbols, we can get two selection rules:  (i) the first (diagonal) term vanishes if $J_a<2$ or if $J_b<2$ and (ii) the second (off-diagonal) term vanishes if $|J_a+J_b|<2$.  Thus the Hg($^3P_0$)+Hg($^3P_2$) collision possesses a long-range $C_5/R^5$ interaction even though the Hg($^3P_0$) atom has no quadrupole moment.  Using the conditions $M_a+M_b = M_a^{\prime}+M_b^{\prime}= \Omega$, where $\Omega$ is the total angular momentum projection and a good molecular quantum number,  we get the matrix elements
{\setlength{\arraycolsep}{2mm}
\begin{eqnarray} \label{eq:H2}
&&\hspace*{-0.3in}H_{M_a,M_a^{\prime}}^{(S_a,S_b,J_a,J_b,\Omega)} =  (-1)^{\Omega}R^{-5}  \nonumber  \\ 
&&\times  \left[ \frac{24}{(2-M_a+M_a^{\prime})! (2+M_a-M_a^{\prime})! } (-1)^{J_a+J_b} \right. \nonumber  \\
&& \left(\! \begin{array}{ccc}J_a & 2 & J_a \\ -M_a & M_a-M_a^{\prime} & M_a^{\prime} \end{array} \!\right)
\left(\! \begin{array}{ccc}J_b & 2 & J_b \\ M_a-\Omega & -(M_a-M_a^{\prime}) & \Omega-M_a^{\prime} \end{array} \!\right) 
 \langle a||Q||a \rangle  \langle b||Q||b \rangle \nonumber \\ 
&& \pm (1-\delta_{a,b}) \frac{24}{(2-M_a-M_a^{\prime}+\Omega)! (2+M_a+M_a^{\prime}-\Omega)! }  \nonumber  \\
&& \left.  \left(\! \begin{array}{ccc}J_a & 2 & J_b \\ -M_a & M_a+M_a^{\prime}-\Omega & \Omega-M_a^{\prime} \end{array} \!\right)
\left(\! \begin{array}{ccc}J_b & 2 & J_a \\ M_a-\Omega & -(M_a+M_a^{\prime}-\Omega) & M_a^{\prime} \end{array} \!\right) 
|\langle a||Q||b \rangle |^2  \right] \ \ \ \ \ \ 
\end{eqnarray}
\!}
where the dimensions of the matrix are given by
$\max(\Omega- J_b, -J_a) \leq M_a \leq \min(\Omega + J_b, J_a)$
and
$\max(\Omega - J_b, -J_a) \leq M_a^{\prime} \leq  \min(\Omega + J_b, J_a)$.

The $C_5$ coefficients are given by the eigenvalues of this matrix.  The eigenvectors can be used to complete identification of the molecular state ${\cal S}=\{S_a,J_a,S_b,J_b,\Omega,\pi,{\cal R},i\}$ where $0\leq\Omega\leq 4$, $\pi=g \text{ or } u$, ${\cal R}=+ \text{ or } -$, and $i$ distinguishes states of the same symmetry.  In a few cases of  double zero eigenvalues it was necessary to take the symmetric and antisymmetric linear combinations of the degenerate eigenvectors.

The resulting $C_5$ coefficients are given in Table \ref{table:C5+C6}.  It can be seen that of the total 90 distinct molecular states, 37 have positive (repulsive) $C_5$ coefficients, 32 have negative (attractive) $C_5$ coefficients, and 21 have zero $C_5$ coefficients.   Of the 21 states with zero $C_5$ coefficients, 12 are ``nonobligatory'' in the sense that nonzero values would be allowed by the selection rules.

\section{Polarizabilities of atoms and molecular $C_6$ coefficients}\label{sec:polarizabilities}
The dispersive van der Waals interaction between two atoms, $C_6/R^6$ can be accurately calculated in terms of the dynamic polarizabilities of the two atoms.  It can be approximately calculated in terms of the static polarizabilities of the two atoms, $\alpha_a$ and $\alpha_b$.  Only the latter are available for excited mercury atoms.  The most frequently used approximation \cite{miller77} for the $C_6$ coefficient in terms of the static polarizabilities is the Slater-Kirkwood formula \cite{slater32}
\begin{equation}\label{eq:SK}
C_6 = \frac{3}{2}\frac{\alpha_a\alpha_b}{\sqrt{\alpha_a/n_a}+\sqrt{\alpha_b/n_b}}
\end{equation}
where $n_a$ and $n_b$ are the numbers of electrons in the outer shells of the two atoms.  However, there exists relatively little experience with excited states.  To validate the approximation and get some idea of its accuracy we sought similar systems where {\it both} accurate static polarizabilites and accurate van der Waals coefficients are available.  Two apropos systems were found: (1) rare-gas dimers with both atoms in excited $^3P_2$ states and (2) zinc dimers with one atom excited ($4^3P$ or $4^1P$) and the other atom in the ground state.  In both cases the $C_6$ coefficients were calculated using the exact Casimir-Polder formula \cite{casimir48} in terms of the frequency-dependent polarizabilities.  The comparison of the Slater-Kirkwood and Casimir-Polder results in Table \ref{table: Slater-Kirkwood} suggests an accuracy of $\sim 20\%$.

The {\it ab initio} values obtained by Rosenkrantz {\it et al.}\ \cite{rosenkrantz80} for the static polarizabilities of the mercury atom in its $6s6p$ $^3P$ and $^1P$ states with $\Sigma(M_L=0)$ and $\Pi(M_L=1)$ projections, neglecting spin-orbit coupling, are given in Table \ref{table: polarizabilities}.  These values have been transformed to the $J$-$M_J$ representation using the transformation matrix given in their paper, which was determined by fitting the experimental atomic term values \cite{mies78}.  This transformation assumes that the radial functions of states with the same $L,S$ but different $J,M_J$ are essentially the same.  No experimental values are available for the excited states, but their values for the ground-state of mercury, $\alpha=36.2$ calculated with a 2-electron relativistic effective core potential (ECP) and 34.3 calculated with a 12-electron relativistic ECP, can be compared with the experimental value, $\alpha=33.91\pm 0.34$ \cite{goebel96a}.  The excited-state polarizabilities that we use for the triplet states were calculated with the 12-electron relativistic ECP, but the singlet excited states were calculated only with the 2-electron relativistic ECP.

Rosenkrantz {\it et al.}\ \cite{rosenkrantz80} also did a similar calculation of the static polarizabilities of the excited zinc atom with a relativistic 2-electron ECP.  We may get some additional idea of the accuracy of the mercury polarizabilities by comparing these results with those of Ellingsen {\it et al.}\ \cite{ellingsen01} for zinc, given in Table \ref{table: Slater-Kirkwood}.  The results of Ellingsen {\it et al.}\ presumably should be of of higher accuracy since they account for core polarization;  also, their value for the ground-state polarizability is in excellent agreement with the experimental value.  This comparison is shown in Table \ref{table:Rosenkrantz-Ellingsen}.  The triplet and singlet excited-state polarizabilties of Rosenkrantz {\it et al.}\ are $\sim 11\%$ and $\sim 78\%$ larger, respectively, than those of Ellingsen {\it et al.}.  This agreement is quite satisfactory for the triplet states, but less so for the singlet states.

These comparisons suggest that the uncertainty in the $C_6$ coefficients, which we present below, arises more from the static polarizabilities themselves than from use of the Slater-Kirkwood approximation.

The eigenvectors $\bf x$ of Eq.\ (\ref{eq:H2}) were used to calculate the polarizabilities for the atomic states forming each molecular state
\begin{equation}
\alpha_i={\bf x}^T({\cal S}){\bf A}{\bf x}({\cal S})
\end{equation}
where $\bf{A}$ is a diagonal matrix containing the polarizabilities of Table \ref{table: polarizabilities}(b) on the diagonal.  The corresponding $C_6$ coefficient is then obtained using Eq.\ (\ref{eq:SK}).  These eigenvectors, which were obtained for the leading asymptotic interaction, are expected to still be a reasonable approximation when higher-order terms are included.  The resulting $C_6$ coefficients are given in Table \ref{table: Slater-Kirkwood}.  They can be seen to vary significantly among the various states, with the interactions involving the singlet atom generally being the larger because of its greater polarizability.  However, the comparison in Table \ref{table:Rosenkrantz-Ellingsen} suggests that the Rosenkrantz {\it et al.}\ polarizabilities for the singlet states may be significantly too large.  If these polarizabilities are reduced by the same factor, 0.56, as for zinc, the $C_6$ coefficients would be reduced by a factor of about 0.66 for the molecular states arising from a $^1P_1$ atom interacting with a $^3P_J$ atom and a factor of about 0.42 for the molecular states arising from the interaction of two $^1P_1$ atoms.  At least for zinc, the uncertainty in the triplet atomic polarizabilities appears to be much smaller---the values of Ellingsen  {\it et al.}\  being only 10\% smaller than those of Rosenkrantz {\it et al.}  Similar changes for mercury would reduce the $C_6$ coefficients by about a factor of 0.86 for molecular states formed from two triplet atoms and by about a factor of 0.92 for molecular states formed from a singlet and a triplet atom.  With both changes simultaneously, the effect on the $C_6$ coefficients for a singlet and triplet pair are approximately multiplicative.

\section{Associative ionization cross sections}\label{sec:xsecs}
The long-range potentials are very important, even dominant, for AI in low-energy collisions of two excited mercury atoms.  For the present estimates of the cross sections, we adopt a model similar to that used in Ref.\ \protect\cite{cohen02}; i.e., ionization is assumed to occur if the trajectory on the quantum-mechanical potential curve has a classical turning point smaller than both the black-sphere distance $R_s$ and the crossing distance $R_x$ (if any) of the potential curve with the Hg$_2^+$ potential curve.  The choice of $R_s=4$ \AA\ is discussed in Ref.\ \cite{cohen02}.  The sensitivity to this choice will be examined below.  The relevant data characterizing the short-range potentials, from Ref.\ \cite{cohen02}, is given in Table \ref{table:crossings}.\footnote{A clarification may be helpful.  In Ref.\ \protect\cite{cohen02} the cross sections ignoring the barriers in the {\it ab initio} potential curves (labeled ``nb'' there) were calculated neglecting the potential only at $R>5$ \AA.  In the present calculations, the short-range potential is taken to be completely characterized by the values of $R_s$, $V_s$, $R_x$, and $V_x$ in Table \protect\ref{table:crossings} (interactions not listed in this table have potential curves too repulsive for the AI reaction to occur).  Also, note two typos in Ref.\ \protect\cite{cohen02}: (1) In Table II, ${^3P_1}+{^3P_0}$ has only one state of $0_g^-$ symmetry, and (2) in the caption of Fig.\ 3, $0_u^-$=dash dot.}  For each potential curve, the AI cross section is taken to be
\begin{equation} \label{eq:xsecs}
\sigma=\pi [\min(b_s,b_x,b_o)]^2 .
\end{equation}
The first two impact parameters depend only on the short-range potential.  The largest impact parameter accessing the black-sphere distance is
\begin{equation}
b_s=\left(1-\frac{V(R_s)}{E}\right)^{1/2}R_s
\end{equation}
and the largest impact parameter accessing the crossing into the continuum is
\begin{equation}
b_x=\left(1-\frac{V(R_x)}{E}\right)^{1/2}R_x ,
\end{equation}
assuming the collision energy $E$ exceeds the potential energies $V(R)$ at these points.
The value of $b_o$, determined by the long-range potential, is given by [see Appendix]
\begin{equation}
b_o=\left( \frac{R_o^3}{2E} \frac{dV(R_o)}{dR} \right)^{1/2}
\end{equation}
where $R_o$ is obtained by numerically solving
\begin{equation}
\frac{R_o}{2} \frac{dV(R_o)}{dR} + V(R_o) = E ,
\end{equation}
with
\begin{equation}
V(R)=\frac{C_5}{R^5}+\frac{C_6}{R^6} ,
\end{equation}
for its largest real root.

It can be seen in Table \ref{table:crossings} that most of the contributions to the AI cross sections at a collision energy of 0.00095 a.u.\ are limited by the long-range potentials, which determine the classical orbiting impact parameter.  However, it should be kept in mind that the potential energy at the curve crossing (if any) and the black-sphere distance provide a necessary condition.  For AI to occur, the collision energy must exceed the potential energy at these distances.  Thus, without tunneling, which can be expected to be negligible except possibly in ultracold collisions, the cross section will vanish at sufficiently low energies if the interatomic potential energy is positive at either of these distances.  At the collision energy of 0.00095 a.u., 47 of the 90 potential curves do not contribute to AI for this reason.  At still lower collision energies, the contributions that are energetically allowed are increasingly determined by the long-range potentials.  At $E\lesssim 2\times 10^{-5}$ a.u., the non-zero contributions are entirely determined by the long-range potentials.

We now examine the sensitivity of the cross sections to the parameters, in particular $C_6$ and $R_s$, which are most uncertain.  The classical orbiting cross section is simply obtained for pure power-law potential energy curves [see Appendix]: for $V(R)=C_5/R^5$ it is
$\sigma_{{\rm orb}(5)}=\frac{5\pi}{3}\left(\frac{-3C_5}{2E}\right)^{2/5}$
and for $V(R)=C_6/R^6$,
$\sigma_{{\rm orb}(6)}=3\pi \left(\frac{-C_6}{4E}\right)^{1/3}$.
With the dependence $C_6^{1/3}$, a reduction in $C_6$ by a factor of 0.66 would decrease the cross section by only a factor of 0.87, so the sensitivity is not too great.  The uncertainty due to the parameters associated with the short-range potentials, in particular whether $V_s$ and $V_x$ are positive or negative, is greater than this.

For low-energy collisions, there is little uncertainty due to $R_S$, which is more-or-less arbitrarily chosen (see Ref.\ \cite{cohen02}).  This does not necessarily mean that the magnitude of the autoionization width, which is presently unknown, is unimportant.  It does mean that, accepting the black-sphere model, the value taken for $R_s$ is not critical.  If the width is large enough to saturate ionization in the energetically allowed region, its exact magnitude will be unimportant.  If not saturated, the width will reduce the size of the cross section by a factor that approaches a constant at low energies.

The total AI cross sections are given by statistically weighted sums over all the potential curves arising from the pair of atomic reactants $i\equiv\ ^{2S+1}P_{J}$ and $i^{\prime}\equiv\ ^{2S^{\prime}+1}P_{J^{\prime}}$,
\begin{equation} \label{eq:statwt}
\sigma_{i+i^{\prime}}=(2J_i+1)^{-1}(2J_{i^{\prime}}+1)^{-1}(2-\delta_{i,i^{\prime}})^{-1}\sum_{\Omega_\pi^r} (2-\delta_{\Omega,0})\sigma_{i+i^{\prime}}^{(\Omega_\pi^r)} ,
\end{equation}
where the sum goes over the molecular states listed in Table \ref{table:C5+C6} for each reaction.  This sum reflects the double degeneracy of the $\Omega\neq 0$ states and the pairing of $g$--$u$ states in the cases of
nonidentical states.

\begin{figure}
\scalebox{0.75}{\includegraphics{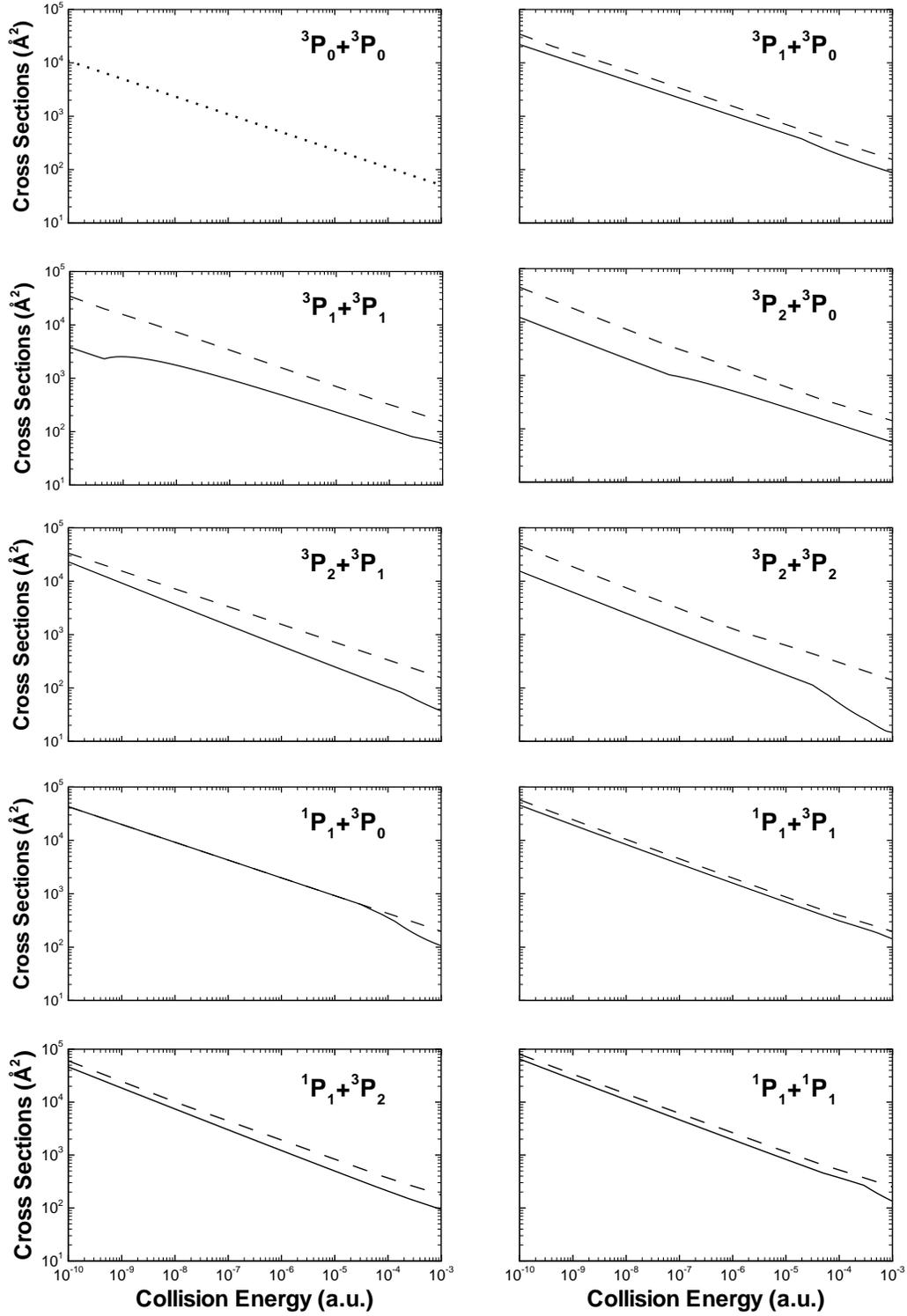}}
\caption{Associative ionization cross sections for collisions of two mercury atoms in the indicated states. The solid curves take into account the long-range and short-range potentials (Eq.\ (\protect\ref{eq:xsecs})).  The dashed curves take into account only the long-range potentials (Eq.\ (\protect\ref{eq:xsecsLR})).  The cross section for Hg($^3P_0$)+Hg($^3P_0$) is shown dotted as a special case (see text).}
\label{fig:xsecs}
\end{figure}

The resulting cross sections are shown in Fig.\ \ref{fig:xsecs}.  These cross sections are expected to be valid down to energies where the scattering is predominantly $s$-wave.  The maximum contributing angular momentum is roughly
\begin{equation}
L_{\text{max}}=(2\mu E)^{1/2}b_o ,
 \end{equation}
 where $\mu$ is the reduced mass and the impact parameter $b_o$ is given in the Appendix.  For an attractive $C_5/R^5$ potential, $L_{\rm max} \approx 847(-C_5)^{1/5}E^{3/10}$, and for a $C_6/R^6$ potential, $L_{\rm max} \approx 831(-C_6)^{1/6}E^{1/3}$ for Hg$^*$+Hg$^*$ collisions.
Thus the semiclassical description, which requires $L_{\rm max} \gg 1$, should be valid above collision energies of $\sim 5\times 10^{-12}$ a.u.\ (temperature of $\sim 2$  $\mu$K) with typical $C_5$ coefficients and above $\sim 5\times 10^{-11}$ a.u.\ (temperature of $\sim 20$  $\mu$K) when all $C_5$ coefficients vanish.

Given the uncertainty in the short-range potentials, it is of interest to consider another approximation to the AI cross sections that might have validity at very low collision energies.  Such an approximation can be obtained by considering only the long-range potentials, i.e.
\begin{equation} \label{eq:xsecsLR}
\sigma_{\rm LR}=\pi b_o^2
\end{equation}
instead of Eq.\ \ref{eq:xsecs}.
This approximation neglects any reduction due to repulsion in the short-range potential.  These cross sections are shown as dashed curves in Fig.\ \ref{fig:xsecs}.

At very low energies the dependence of the cross section on energy tends to be a power law: $E^{-2/5}$ in cases with contributing attractive $C_5/R^5$ potentials and $E^{-1/3}$ otherwise.  At sufficiently low energies, potential curves with positive $C_5$ coefficients will cease to contribute to the cross section, and potential curves with negative $C_5$ coefficients will have this contribution dominate over any contribution from the $C_6$ potential.  The thermal average
\begin{equation}
\bar{\sigma}(T) = \frac{1}{(kT)^2}\int_0^\infty \sigma(E) E e^{-E/kT} dE
 \end{equation}
is particularly easy to obtain for a cross section having power law dependence.  For
\begin{equation}
\sigma(E) = c E^n ,
\end{equation}
\begin{equation}
\bar{\sigma}(T) = \Gamma(n+2)\sigma(kT).
\end{equation}
where $\Gamma$ is the Euler gamma function, with pertinent values $\Gamma(-\frac{1}{3}+2)=0.9027$ and $\Gamma(-\frac{2}{5}+2)=0.8935.$

The cross section for AI in the ${^3P_0}+{^3P_0}$ collision is displayed dotted because it is a special case, subject to additional uncertainty.  This cross section is presented as an upper bound, as are all the black-sphere cross sections, but, unlike the other cases, the true cross section is expected to be {\it considerably smaller}.  In order for this reaction to occur at all in cold collisions, the Hg$(^3P_0)$+Hg$(^3P_0)$ asymptote must be above the minimum in the Hg$_2^+$ well.  Though not all previous work has agreed on the sufficiency of the binding energy of Hg$_2^+$, the most reliable experimental and theoretical results now agree that this well depth is greater than the required 1.10 eV (see Ref.\ \cite{cohen02} for discussion of this point).  However, the single potential curve coming from the Hg($^3P_0)$+Hg$(^3P_0)$ asymptote does not cross the Hg$_2^+$ curve, so adiabatically AI still would not occur.  The reason AI is nonetheless possible is that there is an avoided crossing of this curve with another curve of $0_g^+$ symmetry coming from the Hg($^3P_1)$+Hg$(^3P_1)$ asymptote.  Their diabatic crossing probability is $\sim 0.33$, only weakly dependent on the collision energy as long as $E$ is small compared with the potential energy of 0.013 a.u.\ at the crossing.  An important point here is that the state corresponding to this potential curve has 93\% quintet character and quintet states do not couple to the ionization continuum (only singlet and triplet states can be formed by coupling Hg$_2^+$ to an electron, and the relevant coupling matrix element conserves spin).  Thus the autoionization width is reduced by a factor of $\sim 14$.  Even if the black-body model is valid for the other reactions, it is much less likely to be valid with an autoionization width this small.

\section{Conclusions}\label{sec: conclusions}
The uncertainty of long-range barriers, which affected the results in Ref.\ \cite{cohen02}, has been resolved.  The long-range $1/R^5$ interactions, which may be attractive or repulsive, have been accurately calculated.  The $1/R^6$ interactions, which are generally attractive, have been determined less accurately, but still sufficient for useful estimates of the AI cross sections.

The only AI cross section for mercury that has been experimentally measured is for Hg($^3P_1)$+Hg$(^3P_0)$.  The most reliable values are 99 \AA$^2$ and 64 \AA$^2$, with error bars of 25 \AA$^2$, which come from two different analyses of the same experiment \cite{majetich91} (see footnote 5 of Ref.\ \cite{cohen02}), although there is an older published experimental value \cite{tan68} of 460 \AA$^2$.  The present thermally averaged value for temperature 300 K is 83 \AA$^2$, in agreement with the most recent experiment.  This agreement tends to validate the black-sphere model, which, in principle, provides an upper limit on the cross sections.  Another experimental value \cite{sepman84} of $160\pm 40$ \AA$^2$ has been published for the Hg($^3P_0)$+Hg$(^3P_0)$ reaction, but has been generally repudiated on experimental grounds \cite{majetich91,majetich89,sibata79} and is inconsistent with the present upper limit on the cross section for this reaction.

The cross section for AI in the Hg$(^3P_0)$+Hg$(^3P_0)$ deserves attention because of the proposed use of $^3P_0$ as an atomic clock state, even if the discussion is necessarily speculative.  Compared with the other reaction pairs, the ``strength'' of this reaction may be reduced by three factors: a factor of $\sim 0.33$ due to the required curve hopping, a factor of $\sim 0.07$ due to the non-ionizing quintet component, and a factor due to  the time spent in the continuum being fairly short.  Only the first factor has been taken into account in the cross section displayed as a dotted line in Fig.\ \ref{fig:xsecs}.  The other two factors could easily reduce the cross section by another order of magnitude or more.  Better theoretical estimates will require determination of the relevant autoionization widths.

\begin{acknowledgments}
This work was done under the auspices of the U.S. Department of Energy.
The work of A.D. was supported in part by the National
Science Foundation.
\end{acknowledgments}

\appendix*
\section{Classical orbiting cross sections}
The classical orbiting distance $R_o$ and impact parameter $b_o$ at collision energy $E$ are given by the solutions of the simultaneous equations,
\begin{subequations}
\label{eq:ClOrb}
\begin{equation}
V_{\rm{eff}}(R)=E
\end{equation}
and
\begin{equation}
\frac{dV_{\rm{eff}}(R)}{dR}=0 ,
\end{equation}
\end{subequations}\\
where
\begin{equation}
V_{\rm{eff}}(R)=V(R)+\frac{b^2E}{R^2} ,
\end{equation}
for $R$ and $b$.
The corresponding cross section is
\begin{equation}
\sigma_{{\rm orb}}=\pi b_o^2 .
\end{equation}
For
\begin{equation}
V(R)= \frac{C_n}{R^n},
\end{equation}
\begin{equation}
\sigma_{{\rm orb}(n)}=\frac{\pi n}{n-2}\left(\frac{-(n-2)C_n}{2E}\right)^{2/n} ,
\end{equation}
so
$\sigma_{{\rm orb}(5)}=\frac{5\pi}{3}\left(\frac{-3C_5}{2E}\right)^{2/5}$
and
$\sigma_{{\rm orb}(6)}=3\pi \left(\frac{-C_6}{4E}\right)^{1/3}$.
For
\begin{equation}
V(R)= \frac{C_5}{R^5} + \frac{C_6}{R^6} ,
\end{equation}
the Eqs.\ (\ref{eq:ClOrb}) can be solved numerically.  The largest positive real root, assuming such exists, gives the desired result; otherwise, classical orbiting does not occur.

\newpage
\begin{table}
\caption{Quadrupole matrix elements of the excited mercury atom (Hg $6s6p$) in atomic units. }
\label{table: quadrupole}
\begin{ruledtabular}
\begin{tabular}{ccccc}
\multicolumn{5}{l}{(a) Reduced matrix elements.}\\
& $^3P_0$ &  $^3P_1$ & $^3P_2$ & $^1P_1$ \\
\cline{2-5}
$^3P_0$ &  0 & 0 & $-10.22$ & 0\\
$^3P_1$ & 0 & $8.218$ & $-15.78$ & $3.131$  \\
$^3P_2$ &  $-10.22$ &  $15.78$ & $-15.38$ & $2.860$ \\
$^1P_1$ &  0 &  $3.131$ & $-2.860$ & $-29.67$ \\
\hline
\multicolumn{5}{l}{(b) Conventional quadrupole moments.}\\
& $^3P_0$ &  $^3P_1$ & $^3P_2$ & $^1P_1$ \\
\cline{2-5}
& 0 & 3.00 & $-7.35$ & $-10.83$ \\
\end{tabular}
\end{ruledtabular}
\end{table}

\clearpage
%\begin{table}
\begin{longtable}[p]{@{\extracolsep{0.2in}}cccccc}
\caption{Long-range interaction potentials for Hg($6s6p$)+Hg($6s6p$) with spin-orbit coupling (Hund's case c),  in atomic units. } \\
%\begin{ruledtabular}
\hline\hline

\multicolumn{6}{l}{\underline{$^3P_2$+$^3P_2$}} \\
$0_g^+$  & $0_g^+$ & $0_g^+$ & $0_u^-$ & $0_u^-$ & $1_g$ \\
$+\frac{154.3}{R^5}      -\frac{704}{R^6}$ &  
$+\frac{71.4}{R^5}        -\frac{1074}{R^6}$ & 
$-\frac{22.9}{R^5}         -\frac{768}{R^6}$ & 
$+\frac{131.2}{R^5}      -\frac{600}{R^6}$ & 
$-\frac{50.1}{R^5}         -\frac{816}{R^6}$ & 
$-\frac{66.8}{R^5}         -\frac{777}{R^6}$  \\  
\hline
$1_g$ & $1_u$ & $1_u$ & $2_g$ & $2_g$ & $2_u$ \\
$+\frac{53.3}{R^5}       -\frac{971}{R^6}$ & 
$+\frac{56.9}{R^5}       -\frac{1040}{R^6}$ & 
$-\frac{43.4}{R^5}        -\frac{714}{R^6}$ & 
$-\frac{87.9}{R^5}        -\frac{814}{R^6}$ & 
$+\frac{40.6}{R^5}       -\frac{932}{R^6}$ & 
$-\frac{94.6}{R^5}        -\frac{787}{R^6}$ \\ 
\hline
$3_g $ & $3_u $ & $4_g $ \\
$-\frac{121.7}{R^5}      -\frac{705}{R^6}$ & 
$+\frac{40.6}{R^5}       -\frac{705}{R^6}$ & 
$+\frac{81.1}{R^5}       -\frac{478}{R^6}$  \\ 
 \hline

\multicolumn{6}{l}{\underline{$^3P_2$+$^3P_1$}} \\
$0_{g}^+$  & $0_{g}^-$  & $0_{g}^-$  & $0_{u}^+$  & $0_{u}^-$  & $0_{u}^-$  \\ 
$+\frac{199.1}{R^5}       -\frac{972}{R^6}$ & 
$-\frac{212.9}{R^5}       -\frac{932}{R^6}$ & 
$+\frac{13.9}{R^5}      -\frac{893}{R^6}$ & 
$-\frac{132.9}{R^5}        -\frac{972}{R^6}$ & 
$+\frac{146.7}{R^5}       -\frac{966}{R^6}$ & 
$-\frac{80.0}{R^5}        -\frac{849}{R^6}$ \\
\hline
$1_{g}$  & $1_{g}$  & $1_{g}$  & $1_{u}$  & $1_{u}$  & $1_{u}$ \\
 $-\frac{132.7}{R^5}      -\frac{896}{R^6}$ & 
 $-\frac{53.7}{R^5}      -\frac{809}{R^6}$ & 
 $+\frac{20.6}{R^5}           -\frac{827}{R^6}$ &  
 $+\frac{188.0}{R^5}       -\frac{971}{R^6}$ & 
 $-\frac{107.6}{R^5}      -\frac{750}{R^6}$ & 
 $+\frac{19.3}{R^5}       -\frac{820}{R^6}$  \\ 
  \hline
$2_{g}$ & $2_{g}$ & $2_{u}$ & $2_{u}$ & $3_{g}$ & $3_{u}$ \\
$+\frac{232.2}{R^5}      -\frac{822}{R^6}$ & 
$+\frac{33.2}{R^5}      -\frac{678}{R^6}$ & 
$-\frac{132.9}{R^5}     -\frac{972}{R^6}$ & 
$+\frac{33.0}{R^5}     -\frac{544}{R^6}$ & 
$-\frac{99.5}{R^5}     -\frac{681}{R^6}$ & 
$+\frac{33.3}{R^5}        -\frac{681}{R^6}$  \\  
\hline

\multicolumn{6}{l}{\underline{$^3P_2$+$^3P_0$}} \\
$0_{g}^+$  & $0_{u}^+$  & $1_{g}$ & $1_{u}$  & $2_{g}$ & $2_{u}$  \\
$+\frac{125.3}{R^5}     -\frac{947}{R^6}$ &  
$-\frac{125.3}{R^5}      -\frac{947}{R^6}$ & 
$-\frac{83.6}{R^5}      -\frac{869}{R^6}$ & 
$+\frac{83.6}{R^5}     -\frac{869}{R^6}$ &  
$+\frac{20.9}{R^5}     -\frac{611}{R^6}$ &  
$-\frac{20.9}{R^5}      -\frac{611}{R^6}$ \\  
\hline

\multicolumn{6}{l}{\underline{$^3P_1$+$^3P_1$}} \\
$0_g^+$  & $0_g^+$ & $0_u^-$ & $1_g$ & $1_u$ & $2_g$ \\
$+\frac{81.0}{R^5}      -\frac{736}{R^6}$ & 
$                                    -\frac{857}{R^6}$ & 
$                                    -\frac{984}{R^6}$ & 
$-\frac{54.0}{R^5}      -\frac{796}{R^6}$ & 
$                                    -\frac{796}{R^6}$ & 
$+\frac{13.5}{R^5}     -\frac{984}{R^6}$  \\ 
\hline

\multicolumn{6}{l}{\underline{$^3P_1$+$^3P_0$}} \\
$0_{g}^-$  & $0_{u}^-$  & $1_{g}$ & $1_{u}$  \\
$-\frac{699}{R^6}$ & 
$-\frac{699}{R^6}$ & 
$-\frac{879}{R^6}$ & 
$-\frac{879}{R^6}$   \\   
\hline

\multicolumn{6}{l}{\underline{$^3P_0$+$^3P_0$}} \\
$0_g^+$  & \\
$-\frac{787}{R^6}$ &  \\   
\hline

\newpage

\multicolumn{6}{l}{\underline{$^1P_1$+$^3P_2$}} \\
$0_{g}^+$  & $0_{g}^-$  & $0_{g}^-$  & $0_{u}^+$  & $0_{u}^-$  & $0_{u}^-$   \\ 
$-\frac{114.0}{R^5}    -\frac{1401}{R^6}$ & 
$+\frac{271.6}{R^5}     -\frac{3309}{R^6}$ & 
$-\frac{37.0}{R^5}      -\frac{1655}{R^6}$ & 
$-\frac{124.9}{R^5}    -\frac{1401}{R^6}$ & 
$+\frac{276.0}{R^5}     -\frac{3282}{R^6}$ & 
$-\frac{32.6}{R^5}      -\frac{1688}{R^6}$  \\  
\hline
$1_{g}$  & $1_{g}$  & $1_{g}$ & $1_{u}$  & $1_{u}$  & $1_{u}$  \\
 $+\frac{289.1}{R^5}    -\frac{2032}{R^6}$ & 
 $-\frac{179.5}{R^5}   -\frac{1812}{R^6}$ & 
 $+\frac{5.5}{R^5}         -\frac{1917}{R^6}$ & 
 $+\frac{288.9}{R^5}    -\frac{2044}{R^6}$ & 
 $-\frac{170.8}{R^5}   -\frac{1821}{R^6}$ & 
 $+\frac{5.7}{R^5}         -\frac{1896}{R^6}$ \\    
 \hline
$2_{g}$ & $2_{g}$ & $2_{u}$ & $2_{u}$ & $3_{g}$ & $3_{u}$ \\
$-\frac{337.2}{R^5}    -\frac{2124}{R^6}$ & 
$+\frac{44.5}{R^5}       -\frac{1800}{R^6}$ & 
$-\frac{344.1}{R^5}    -\frac{2121}{R^6}$ & 
$+\frac{39.4}{R^5}       -\frac{1813}{R^6}$ & 
$+\frac{117.3}{R^5}     -\frac{967}{R^6}$ & 
$+\frac{121.7}{R^5}     -\frac{967}{R^6}$  \\   
\hline

\multicolumn{6}{l}{\underline{$^1P_1$+$^3P_1$}} \\
$0_{g}^+$  & $0_{g}^+$  & $0_{g}^-$  & $0_{u}^+$  & $0_{u}^+$  & $0_{u}^-$  \\
$-\frac{280.8}{R^5}    -\frac{2304}{R^6}$ & 
$                                    -\frac{1969}{R^6}$ & 
$                                    -\frac{1419}{R^6}$ & 
$-\frac{304.4}{R^5}     -\frac{2304}{R^6}$ & 
$                                     -\frac{1969}{R^6}$ & 
$                                     -\frac{1419}{R^6}$  \\ 
\hline
$1_{g}$  & $1_{g}$  & $1_{u}$  & $1_{u}$  & $2_{g}$ & $2_{u}$  \\
$+\frac{187.2}{R^5}    -\frac{2159}{R^6}$ & 
$                                     -\frac{2159}{R^6}$ & 
$+\frac{202.9}{R^5}    -\frac{2159}{R^6}$ & 
$                                     -\frac{2159}{R^6}$ & 
$-\frac{46.8}{R^5}       -\frac{1419}{R^6}$ & 
$-\frac{50.7}{R^5}       -\frac{1419}{R^6}$  \\   
\hline

\multicolumn{6}{l}{\underline{$^1P_1$+$^3P_0$}} \\
$0_{g}^-$  & $0_{u}^-$ & $1_{g}$ & $1_{u}$  \\
$-\frac{2843}{R^6}$ &  
$-\frac{2843}{R^6}$ & 
$-\frac{1262}{R^6}$ & 
$-\frac{1262}{R^6}$   \\   
\hline

\multicolumn{6}{l}{\underline{$^1P_1$+$^1P_1$}} \\
$0_g^+$  & $0_g^+$ & 
$0_u^-$ & $1_g$ & $1_u$ & $2_g$ \\
$+\frac{1056.4}{R^5}    -\frac{8477}{R^6}$ & 
$                                       -\frac{4920}{R^6}$ & 
$                                       -\frac{2077}{R^6}$ & 
$-\frac{704.2}{R^5}       -\frac{6618}{R^6}$ & 
$                                       -\frac{6618}{R^6}$ & 
$+\frac{176.1}{R^5}      -\frac{2077}{R^6}$ \\

%\end{ruledtabular}
\hline\hline
%\tablenotetext[1]{}
\label{table:C5+C6}
\end{longtable}
%\end{table}

\clearpage

\begin{table}
\caption{Comparison of $C_6$ coefficients calculated by the exact Casimir-Polder (CP) formula using the dynamic polarizabilities with those approximated by the Slater-Kirkwood (SK) formula (Eq.\ \protect\ref{eq:SK}) using only the static polarizabilities $\alpha$.  These results are presented to validate the SK formula for excited states.}
\label{table: Slater-Kirkwood}
\begin{ruledtabular}
\begin{tabular}{lcccc}
\multicolumn{4}{l}{(a) Rare-gas dimers ($n_a=n_b=1$ in the SK formula).}\\
   & $\alpha_a$$^a$  & $\alpha_b$$^a$ & \multicolumn{2}{c}{$C_6$}\\
\cline{4-5}
&   & & CP$^b$ & SK$$\\
\cline{2-5}
Ne$(3s,\ ^3P_{2,2})$+Ne$(3s,\ ^3P_{2,2}), \ 4g$ & 180.2 & 180.2 & 1877 & 1814\\
Ar$(4s,\ ^3P_{2,2})$+Ar$(4s,\ ^3P_{2,2}), \ 4g$ & 301.7 & 301.7 & 4417 & 3930\\
Kr$(5s,\ ^3P_{2,2})$+Kr$(5s,\ ^3P_{2,2}), \ 4g$ & 315.8 & 315.8 & 4994 & 4209\\
Xe$(6s,\ ^3P_{2,2})$+Xe$(6s,\ ^3P_{2,2}), \ 4g$ & 387.4 & 387.4 & 7138 & 5719\\
\hline
\multicolumn{4}{l}{(b) Zinc dimers ($n_a=n_b=2$ in the SK formula).}\\
&   $\alpha_a$$^c$ &    $\alpha_b$$^c$ & \multicolumn{2}{c}{$C_6$} \\
\cline{4-5}
 &  & & CP$^c$ & SK$$\\
 \cline{2-5}
Zn$(4s^2,\ ^1S)$+Zn$(4s^2,\ ^1S)$, $^1\Sigma$ & 39.12 & 39.12  & 282    & 260 \\
Zn$(4s^2,\ ^1S)$+Zn$(4s^2,\ ^3P)$, $^3\Sigma$ & 39.12 & 89.08  & 435    & 471 \\
Zn$(4s^2,\ ^1S)$+Zn$(4s^2,\ ^3P)$, $^3\Pi$ & 39.12 & 55.21 & 370 & 335 \\
Zn$(4s^2,\ ^1S)$+Zn$(4s^2,\ ^1P)$, $^1\Sigma$ & 39.12 & 396.74  & 1139 & 1258   \\
Zn$(4s^2,\ ^1S)$+Zn$(4s^2,\ ^1P)$, $^1\Pi$ & 39.12  & 116.51  & 674 & 567  \\
\end{tabular}
\end{ruledtabular}
\tablenotetext[1]{From Ref.\ \protect\cite{molof74}.}
\tablenotetext[2]{From Ref.\ \protect\cite{derevianko00}; dynamic polarizabilities were adjusted to agree with Ref.\ \protect\cite{molof74} in the static limit.}
\tablenotetext[3]{From Ref.\ \protect\cite{ellingsen01}.}
\end{table}

\begin{table}
\caption{Comparison of static polarizabilities.}
\label{table:Rosenkrantz-Ellingsen}
\begin{ruledtabular}
\begin{tabular}{lccc}
 & \multicolumn{3}{c}{Polarizability} \\
 \cline{2-4}
 & Ellingsen {\it et al.}\ \protect\cite{ellingsen01} & Rosenkrantz {\it et al.}\ \protect\cite{rosenkrantz80} &  Experimental \protect\cite{goebel96b}\\
 \cline{2-4}
Zn$(4s^2,\ ^1S)$ & 39.12  & 35.1 & $38.8\pm 0.8$ \\
Zn$(4s^2,\ ^3P\Sigma)$ & 89.08  & 99.37  \\
Zn$(4s^2,\  ^3P\Pi)$ & 55.21 & 61.37  \\
Zn$(4s^2,\ ^1P\Sigma)$ & 396.74  &  705.45   \\
Zn$(4s^2,\ ^1P\Pi)$ & 116.51  &  207.42  \\
\end{tabular}
\end{ruledtabular}
\end{table}

\begin{table}
\caption{Excited mercury atom (Hg $6s6p$) dipole polarizabilities in atomic units. }
\label{table: polarizabilities}
\begin{ruledtabular}
\begin{tabular}{ccccc}
\multicolumn{5}{l}{(a) {\it Ab initio} calculation in $L$-$S$ representation (from Ref.\ \protect\cite{rosenkrantz80}).}\\
& $\Sigma$ &  $\Pi$ \\
\cline{2-3}
$^3P$ & 128.38 & 58.77 \\
$^1P$ & 532.58 & 158.08 \\
\hline
\multicolumn{4}{l}{(b) Transformed to $J$-$M_J$ representation.}\\
& $M_J=0$ & $M_J=1$ & $M_J=2$ \\
\cline{2-4}
$^3P_0$ & 81.97  \\
$^3P_1$ & 70.03 & 95.11  \\
$^3P_2$ & 105.18 & 93.58 & 58.77 \\
$^1P_1$ & 521.31 & 156.55  \\
\end{tabular}
\end{ruledtabular}
\end{table}

\clearpage

\begingroup
\squeezetable
%\begin{table}
\begin{longtable}[p]{@{\extracolsep{0.2in}}ccccccccccc}
\caption{Distances $R$ (in \AA) and potential energies $V$ (in a.u.) of the black sphere radius (subscript $s$), the crossing with the molecular-ion potential curve if any (subscript $x$), and classical-orbiting (subscript $o$).  The corresponding impact parameters $b$ (in \AA) are given for a collision energy 0.00095 a.u.  Only the potential curves contributing to AI cross sections at this collision energy are listed.  The cross section is determined by the smallest $b$ (shown in bold face).} \\
%\begin{ruledtabular}
\hline\hline
%\begin{tabular}{ccccccccccc}
&& \multicolumn{3}{c}{Black sphere} & \multicolumn{3}{c}{Continuum crossing} & \multicolumn{3}{c}{Classical orbiting} \\
\cline{3-5}  \cline{6-8}  \cline{9-11}
Reaction & Symmetry & $R_s$ & $V_s$ & $b_s$ & $R_x$ & $V_x$ & $b_x$ & $R_o$ & $V_o$ & $b_o$ \\
\hline
$^3P_0$ + $^3P_0$$^{\ a}$ \\
&$0_g^+$&4.00& $-$0.00954& 13.29& 2.97& $-$0.00620& 8.15& 5.76& $-$0.000475& {\bf 7.05}\\
$^3P_1$ + $^3P_0$ \\
&$0_g^-$&4.00& $-$0.0169& 17.33& 2.96& $-$0.0139& 11.70& 5.64& $-$0.000475& {\bf 6.91}\\
&$1_g$&4.00& $-$0.00831& 12.49& 3.20& $-$0.0116& 11.65& 5.86& $-$0.000475& {\bf 7.18}\\
&$0_u^-$&4.00& $-$0.000195& {\bf 4.39}& 3.70& $-$0.000712& 4.89& 5.64& $-$0.000475& 6.91\\
$^3P_1$ + $^3P_1$ \\
    &$0_g^+$&4.00& $-$0.00820& 12.41& 2.97& $-$0.0223& 14.69& 5.84& $-$0.000475& {\bf 7.15}\\
&$1_g$&4.00& $+$0.000269& {\bf 3.39}&4.12&0.000203&3.65& 6.24& $-$0.000534& 7.80\\
&$2_g$&4.00& $-$0.00637& 11.10& 3.64& $-$0.0103& 12.54& 5.86& $-$0.000455&{\bf 7.12}\\
$^3P_2$ + $^3P_0$ \\
&$2_g$&4.00& $-$0.00178& 6.78&&&& 5.26 & $-$0.00421&{\bf 6.31}\\
&$2_u$&4.00& $-$0.00475& 9.79&&&& 5.75 & $-$0.000509&{\bf 7.13}\\
$^3P_2$ + $^3P_1$ \\
&$1_g$&4.00& $+$0.000897& {\bf 0.95}&&&& 6.82 & $-$0.000568 & 8.62\\
&$3_g$&4.00& $-$0.0148& 16.29&&&& 6.47 & $-$0.000566 & {\bf 8.18}\\
&$1_u$&4.00& $-$0.00129& {\bf 6.14}&&&& 6.58 & $-$0.000566 & 8.31\\
&$2_u$&4.00& $-$0.00355& 8.70&&&& 6.87 & $-$0.000565 & {\bf 8.67}\\
$^3P_2$ + $^3P_2$ \\
&$0_g^+$&4.00& $-$0.00134& {\bf 6.22}&&&& 5.95&  $-$0.000507&7.37\\
&$3_g$&4.00& $-$0.000478& {\bf 4.90}&&&& 6.64 & $-$0.000573& 8.40\\
&$4_g$&4.00& $-$0.0222& 19.76&&&& 3.69  & $+$0.000762&{\bf 1.64}\\
&$0_u^-$&4.00& $-$0.000567& {\bf 5.05}&&&& 6.22&  $-$0.000531&7.77\\
$^1P_1$ + $^3P_0$ \\
&$0_g^-$&4.00& $-$0.0174& 17.56&&&& 7.13 & $-$0.000475&{\bf 8.73}\\
&$1_g$&4.00& $-$0.000319& {\bf 4.62}&&&& 6.23 & $-$0.000475&7.63\\
&$0_u^-$&4.00& $-$0.000759& {\bf 5.37}&&&& 7.13&  $-$0.000475&8.73\\
&$1_u$&4.00& $-$0.000800& {\bf 5.43}&&&& 6.23 & $-$0.000475&7.63\\
$^1P_1$ + $^3P_1$ \\
&$0_g^+$&4.00& $-$0.0141& 15.92&&&& 7.95&  $-$0.000567&{\bf 10.05}\\
&$0_g^+$&4.00& $-$0.00522& 10.19&&&& 6.71&  $-$0.000475&{\bf 8.21}\\
&$0_g^-$&4.00& $-$0.00174& {\bf 6.73}&&&& 6.35&  $-$0.000475&7.78\\
&$1_g$&4.00& $-$0.00454& 9.61&&&& 6.81&  $-$0.000475&{\bf 8.34}\\
&$2_g$&4.00& $-$0.00228& {\bf 7.37}&&&& 6.64&  $-$0.000513&8.24\\
&$0_u^+$&4.00& $-$0.0219& 19.62&&&& 8.02&   $-$0.000570&{\bf 10.15}\\
&$0_u^+$&4.00& $-$0.0209& 19.20&&&& 6.71 & $-$0.000475&{\bf 8.21}\\
&$1_u$&4.00& $-$0.00764& 12.02&&&& 6.81&  $-$0.000475&{\bf 8.34}\\
&$1_u$&4.00& $-$0.00513& 10.12&&&& 5.48& $-$0.000048&{\bf 5.61}\\
\newpage
$^1P_1$ + $^3P_2$ \\
&$0_g^+$&4.00& $-$0.0118& 14.67&&&& 6.99 & $-$0.000546 & {\bf  8.78}\\
&$1_g$&4.00& $-$0.0143& 16.00&&&& 7.45 & $-$0.000556  & {\bf 9.38}\\
&$1_g$&4.00& $+$0.000060& {\bf 3.87}&&&& 6.65 & $-$0.000471 & 8.13\\
&$2_g$&4.00& $-$0.0214& 19.39&&&& 8.07 & $-$0.000577 & {\bf 10.23}\\
&$1_u$&4.00& $-$0.0254& 21.05&&&& 7.42 & $-$0.000554 & {\bf 9.34}\\
&$1_u$&4.00& $+$0.000823& {\bf 1.46}&&&& 6.63 &  $-$0.00047 & 8.11\\
&$2_u$&4.00& $-$0.0211& 19.26&&&& 8.09 & $-$0.000578 & {\bf 10.26}\\
&$3_u$&4.00& $-$0.00608& 10.88&&&& 4.53 & $+$0.0001883 & {\bf 4.05}\\
$^1P_1$ + $^1P_1$ \\
&$0_g^+$&4.00& $-$0.0257& 21.18&&&& 7.81&  $-$0.000475&{\bf 9.57}\\
&$0_g^+$&4.00& $-$0.00580& 10.66&&&& 5.33&  $+$0.00207& {\bf  0}\\
&$1_g$&4.00& $-$0.00384& {\bf 8.98}&&&& 9.52 & $-$0.000568&12.04\\
&$2_g$&4.00& $-$0.00820& 12.41&&&& 5.61& $-$0.000147&{\bf 6.03}\\
&$0_u^-$&4.00& $-$0.0157& 16.76&&&& 6.77  &$-$0.000475&{\bf 8.29}\\
%\end{tabular}
%\end{ruledtabular}
\hline\hline
\multicolumn{11}{l}{$^a$With curve crossing taken into account.}
% \tablenotetext[1]{With curve crossing taken into account.}
%\end{table}
\label{table:crossings}
\end{longtable}
\endgroup

\clearpage

\end{document}